\documentclass[twocolumn,preprintnumbers, superscriptaddress, reprint, longbibliography, prx]{revtex4-1}
\usepackage{amsmath}
\usepackage{stmaryrd}
\usepackage{txfonts}
\usepackage{amssymb}
\usepackage{mathrsfs}
\usepackage{graphicx}
\usepackage{dcolumn}
\usepackage{bm}
\usepackage{braket}
\usepackage{epsfig}
\usepackage{color}
\usepackage{ulem}
\usepackage{bibunits}

\usepackage[colorlinks=true, linkcolor=blue, citecolor=blue, urlcolor=blue]{hyperref}
\newcommand{\Nd}{N_\mathrm{d}}
\newcommand{\Ns}{N_\mathrm{site}}

\begin{document}

\title{Stable Higgs mode in anisotropic quantum magnets}
\author{Ying Su}
\affiliation{Theoretical Division, T-4 and CNLS, Los Alamos National Laboratory, Los Alamos, New Mexico 87545, USA}
\author{A. Masaki-Kato}
\affiliation{Computational Condensed Matter Physics Laboratory, RIKEN Cluster for Pioneering Research (CPR), Wako, Saitama 351-0198, Japan}
\affiliation{Computational Materials Science Research Team, RIKEN Center for Computational Science (R-CCS),  Kobe, Hyogo 650-0047, Japan}
\author{Wei Zhu}
\affiliation{Westlake Institution of Advanced Study, Westlake University, Hangzhou 300024, China}
\author{Jian-Xin Zhu}
\affiliation{Theoretical Division, T-4 and CNLS, Los Alamos National Laboratory, Los Alamos, New Mexico 87545, USA}
\author{Yoshitomo Kamiya}
\email{yoshi.kamiya@sjtu.edu.cn}
\affiliation{School of Physics and Astronomy, Shanghai Jiao Tong University
800 Dongchuan Road, Minhang District, Shanghai 200240, China}

\author{Shi-Zeng Lin}
\email{szl@lanl.gov}
\affiliation{Theoretical Division, T-4 and CNLS, Los Alamos National Laboratory, Los Alamos, New Mexico 87545, USA}

\begin{abstract}

  Low-energy excitations associated with the amplitude fluctuation {of an order parameter} in condensed matter systems can mimic the Higgs boson, an elementary particle in the standard model, and are dubbed as Higgs modes. Identifying
  {the condensed-matter Higgs mode is challenging because it is known in many cases to decay}
  rapidly into other low-energy bosonic modes, which renders the Higgs mode invisible. Therefore, it is desirable to find
  {a way}
  to stabilize the Higgs mode, which can offer an insight into the stabilization mechanism of the Higgs mode in condensed matter physics. In quantum magnets,
  {magnetic order caused by spontaneous symmetry breaking supports transverse (magnons) and longitudinal (Higgs modes) fluctuations. When a continuous symmetry is broken, the Goldstone magnon mode generally has a lower excitation energy than the Higgs mode, causing a rapid decay of the latter.}
  In this work, we show that a stable Higgs mode exists in anisotropic quantum magnets near the quantum critical point between the dimerized
  {and magnetically ordered phases.}
  We find
  {that}
  an easy axis anisotropy increases the magnon gap such that the magnon mode is above the Higgs mode near the quantum critical point, and the decay of the Higgs mode into {the} magnon mode is forbidden kinematically. Our results suggest that the anisotropic quantum magnets provide ideal platforms to explore the Higgs physics in condensed {matter systems.}
\end{abstract}

\date{\today}
\maketitle

\section{Introduction}
The Higgs boson in particle physics is modeled by a gauged bosonic condensate, which is responsible for generating mass for other elementary particles. Higgs-like excitations also emerge in condensed matter {systems}
as a consequence of {spontaneous} symmetry breaking. \cite{pekker_amplitude_2015} Examples include charge density wave \cite{yusupov_coherent_2010}, superconductors \cite{PhysRevB.26.4883,PhysRevLett.115.157002,sherman_higgs_2015,doi:10.1146/annurev-conmatphys-031119-050813}, quantum magnets \cite{PhysRevLett.119.067201,jain_higgs_2017} and cold atom condensates in {a} optical lattice \cite{PhysRevLett.109.010401,endres_higgs_2012,gross_quantum_2017}. In these systems, the Higgs mode is the collective amplitude {fluctuation}
of the complex order parameter or vector fields, and is usually gapped.

When a continuous symmetry is broken, there exist gapless Goldstone modes in addition to the massive Higgs mode. Quantum fluctuations therefore
{may induce a}
decay of the Higgs mode into the low-lying Goldstone {modes,} which causes damping of the Higgs mode. The question is whether the Higgs mode remains stable. 
{Podolsky {\it et al}. \cite{PodolskyArovasPRB} addressed this question by using a field theoretical approach and found}
that the imaginary part of the longitudinal susceptibility associated with the Higgs mode diverges at low frequency $\omega$ as $1/\omega$ for two dimensional {(2D)} systems and $\log(1/|\omega|)$ for three dimensional {(3D)} systems, which can obscure the spectral peak of the Higgs mode.  This motivates {the} authors
{in Refs.~\cite{PhysRevLett.110.140401,PhysRevB.88.235108}}
   {to} propose a scalar susceptibility, where a well defined spectral peak corresponding to the Higgs mode appears
   despite of the strong damping. The scalar susceptibility {is argued to be identified}
   in the {Raman spectroscopy.}   
   {The spectral peak of {the} scalar susceptibility is broadened near the quantum phase transition point in 2D, whereas the peak remains sharp in 3D.}
   This is consistent with the intuition that damping of the Higgs mode is stronger in lower dimensions as a result of the quantum fluctuations of the Goldstone modes. {One may then argue that it is necessary} to consider three dimensional systems in order to have a stable Higgs mode. \cite{PhysRevLett.118.147207} 

   In {magnetic}
   materials, continuous symmetry can be lifted by anisotropy. For instance,
   the spin rotation symmetry can be reduced to
   {$U(1)\times Z_2$ symmetry by either an easy plane anisotropy in $XY$-like systems or an easy axis anisotropy in Ising-like systems.}
   The rotation symmetry can also be lifted by an external magnetic field. The reduced symmetry therefore can stabilize the Higgs mode in quantum magnets, as will be discussed below. The magnons carry spin quantum number $S_m=\pm 1$ while the Higgs mode carries spin quantum number $S_h=0$. A Higgs mode with energy $E_h$
   can decay into a pair
   of magnons
   {$(\mathbf{k}_1, \mathbf{k}_2)$}
   with $S_m=\pm 1$ constrained by the energy
   {and momentum}
   conservation
   {laws, also known as the kinematic condition $E_m(\mathbf{k}_1) + E_m(\mathbf{k}_2) = E_h(\mathbf{k}_1 + \mathbf{k}_2)$.}
   {Some or all of magnon branches}
are gapped in magnets with reduced symmetry, which in turn mitigates the decay of the Higgs mode by
{reducing the phase space satisfying}
the kinematic condition
{in a part of, or even entire, region of the Brillouin zone.}
Especially, when a quantum magnet undergoes a continuous quantum phase transition into the quantum paramagnetic state by tuning
{an}
external parameter, such as pressure, the magnitude of the magnetic moment is suppressed continuously
{down}
to zero at the quantum critical point (QCP). The gap of the Higgs mode becomes small near the transition point, and therefore the decay into
magnon modes is suppressed {when the Higgs mode has lower energy than the magnon modes}. Recently, a stable Higgs mode with long lifetime was detected by inelastic neutron scattering measurement in
{the two dimensional quantum magnet}
$\mathrm{C_9H_{18}N_2CuBr_4}$ with an easy axis anisotropy near a quantum critical point. \cite{hong_higgs_2017} The authors constructed an effective spin Hamiltonian for $\mathrm{C_9H_{18}N_2CuBr_4}$, and derived the dispersion relation of the magnon and Higgs modes using the mean-field bond operator approach. The decay of the Higgs mode in $\mathrm{C_9H_{18}N_2CuBr_4}$ was also investigated {recently} using
quantum Monte Carlo simulations \cite{PhysRevLett.122.127201}.

To investigate the role of spin anisotropy on the stability of the Higgs mode, in this work, we study the Higgs mode in an anisotropic bilayer quantum antiferromagnetic Heisenberg model for spin $S=1/2$ with an easy axis anisotropy by employing the bond operator method, field theoretical approach and quantum Monte Carlo simulation. By combining these methods, we can show clearly how the spin anisotropy suppresses the damping of the Higgs mode. The bilayer Heisenberg model is relevant for several quantum magnets including $\mathrm{BaCuSi_2O_6}$ \cite{PhysRevB.55.8357,PhysRevLett.93.087203,sebastian_dimensional_2006} and $\mathrm{Sr_3Ir_2O_7}$ \cite{PhysRevLett.109.157402,PhysRevB.92.024405}. We note that the collective excitations including the Higgs mode in an isotropic model
{have}
been considered in Ref. \cite{PhysRevB.92.245137}. Upon increasing the interlayer antiferromagnetic interaction, the system undergoes a quantum phase transition from
{the N\'{e}el}
order to
{the}
nonmagnetic dimerized phase by forming interlayer
{spin}
singlet.
{Upon reaching the critical point from the magnetically ordered state,}
the magnitude of the moment vanishes and the Higgs mode becomes gapless.
{However,}
because of the easy axis anisotropy, the magnon modes remain gapped. There exists a region where the dispersion of the magnon modes are above the Higgs mode, which prevents the decay of Higgs mode into the magnon modes and therefore the Higgs mode is long lived. We note in passing that the decay of Higgs mode is already forbidden when the magnon gap is larger than half of the Higgs gap. By tuning the magnon gap, the present model allows to investigate the lifetime of the Higgs mode as a function of
{the}
magnon gap. 

In the remainder of the paper, we will employ the mean field bond
{operator}
approach to construct the phase diagram and derive the Higgs and magnon
{dispersion relations.}
Then we will use the field theoretical approach to study the decay of Higgs mode into magnon modes. In the region where such a decay is
{prohibited because of the the kinematic condition,}
   {the Higgs mode becomes the lowest lying mode with
{a}
very sharp spectral peak}.
   {Finally, we will present the results of our quantum Monte Carlo simulation to study the Higgs and magnon modes near the quantum critical point. The paper is then concluded with a summary.}

\begin{figure}
  \begin{center}
    \includegraphics[width=\columnwidth,bb=0 0 307 254]{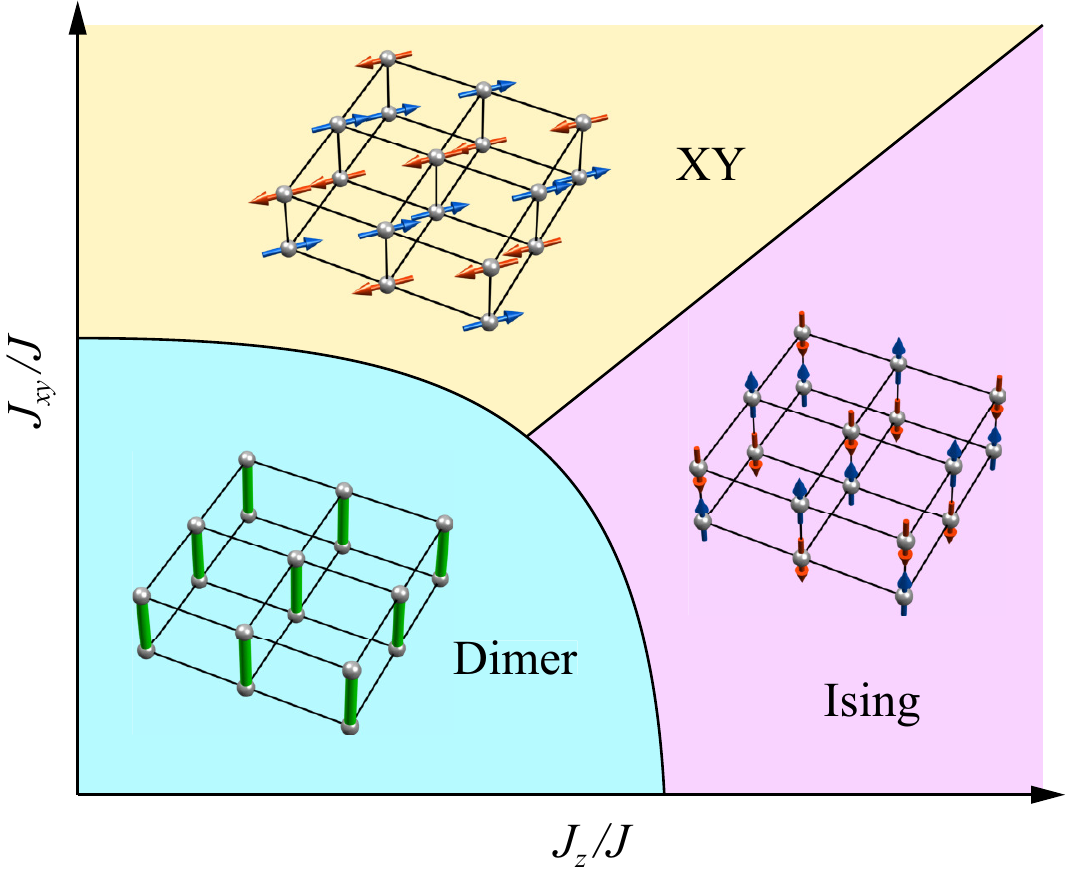}
  \end{center}
  \caption{Schematic phase diagram of the anisotropic quantum magnet model that includes three different phases: interlayer dimer order, Ising and XY AFM order. {The spin texture of the dimer, Ising and \textit{XY} AFM are sketched.}
  } 
  \label{f1}
\end{figure}

\section{Bond operator approach}

We consider
{the}
anisotropic bilayer quantum antiferromagnetic Heisenberg
{(or \textit{XXZ})} model defined on a square lattice. The model Hamiltonian is
\begin{align}
\mathcal{H}=J_{xy}\sum_{l,\langle i j \rangle}[S_{l,i}^x S_{l,j}^x+S_{l,i}^y S_{l,j}^y]+J_z\sum_{l,\langle i j \rangle}[S_{l,i}^z S_{l,j}^z]+J\sum_{i}\mathbf{S}_{1, i}\cdot \mathbf{S}_{2, i}.
\label{eq:H}
\end{align}
where
{$\mathbf{S}_{l,i}$}
is the quantum spin 1/2 operator and $l=1,\ 2$ is the layer index. Here we assume a nearest neighbor anisotropic antiferromagnetic interaction with an Ising-like exchange anisotropy
$J_{z}\ge J_{xy}$
{described by the first two terms} and an antiferromagnetic (AFM) inter-layer coupling {in the last term}.

{In this model, three limits can be identified: (i)}
When $J\gg J_z$ {and $J_{xy}$}, the AFM interlayer coupling stabilizes singlets between aligned spins in different layers. These singlets condense and stabilize a singlet dimer phase. {(ii) For $J_z \gg J$ and $J_{xy}$ in the Ising limit,} 
each layer orders antiferromagnetically with
{spins}
aligned along the $z$ direction and
{staggered}
between layers {that forms the N{\' e}er order}. {(iii) In the $XY$ limit $J_{xy}\gg J$ and $J_z$, {where the spins order antiferromagnetically in the $xy$ plane}. The phase diagram of the model and the corresponding spin structures in the three limits are sketched in Fig. \ref{f1}. Here we focus on the phase boundary between the dimer and AFM phases for $J_z\ge J_{xy}$.} By gradually reducing $J/J_z$, there is a phase transition from
{the} dimer phase to the AFM phase. 
To describe this phase transition, 
{we start with the dimer phase in which spin singlets are stabilized along the vertical bonds between two layers, as shown in Fig. \ref{f1}. The bond operator representation is introduced to describe the dimerized spins by one singlet operator $s_i$ and three triplet operators $t_{i,\alpha}$ with $\alpha=x,\ y,\ z$ as}
\begin{align}
s_{{i}}^\dagger\ket{0}&=\frac{1}{\sqrt{2}}\left(\ket{\uparrow\downarrow}-\ket{\downarrow\uparrow}\right),\\
t_{{i},x}^\dagger\ket{0}&=-\frac{1}{\sqrt{2}}\left(\ket{\uparrow\uparrow}-\ket{\downarrow\downarrow}\right),\\
t_{{i},y}^\dagger\ket{0}&=\frac{i}{\sqrt{2}}\left(\ket{\uparrow\uparrow}+\ket{\downarrow\downarrow}\right),\\
t_{{i},z}^\dagger\ket{0}&=\frac{{1}}{\sqrt{2}}\left(\ket{\uparrow\downarrow}+\ket{\downarrow\uparrow}\right).
\end{align}
We choose $s_i$ and $t_{i,\alpha}$ (here $i$ labels {the interlayer dimers}) to be bosonic operators satisfying the commutation relation
\begin{align}
\left[s_i, s_j^\dagger\right]=\delta_{ij}, \ \  \left[t_{i,\alpha}, t_{j,\beta}^\dagger\right]=\delta_{ij}\delta_{\alpha\beta},\ \  \left[s_{i}, t_{j,\alpha}^\dagger\right]=0.
\end{align}
For each {vertical} bond, it can be either in the singlet or
{one of the triplet states,}
and we have $s_i^\dagger s_i+\sum_{\alpha={x, y, z}}t_{i, \alpha}^\dagger t_{i, \alpha}=1$ for all $i$'s. 
The two spins  $\mathbf{S}_{1,j}$ and $\mathbf{S}_{2,j}$ at the two ends of the $j$-th vertical bond can be expressed in term of the bond operators
\begin{align}
S_{{1,j}}^\alpha &=\frac{1}{2}\left(s_{j}^\dagger t_{{j},\alpha}+t_{{j},\alpha}^\dagger s_{j}-i\epsilon_{\alpha\beta\gamma}t_{{j},\beta}^\dagger t_{{j},\gamma} \right),\\
S_{{2,j}}^\alpha &=\frac{1}{2}\left(-s_{j}^\dagger t_{{j},\alpha}-t_{{j},\alpha}^\dagger s_{j}-i\epsilon_{\alpha\beta\gamma}t_{{j},\beta}^\dagger t_{{j},\gamma} \right),
\end{align}
where $\epsilon_{\alpha\beta\gamma}$ is the Levi-Civita tensor, and
{the}
summation over repeated indices is assumed.

The Hamiltonian $\mathcal{H}$ can be re-expressed in term of these bond operators {as}
\begin{align}
  \mathcal{H}=\frac{J_{xy}}{2}(\mathcal{H}_x+\mathcal{H}_y)+\frac{J_z}{2}\mathcal{H}_z+
  {\frac{J}{4}} \mathcal{H}_J,
\end{align}
\begin{equation}
\begin{split}
    \mathcal{H}_\alpha=& {\sum_{\langle i j \rangle}}\left(s_i^\dagger t_{i,\alpha}+t_{i,\alpha}^\dagger s_i\right)\left(s_j^\dagger t_{j,\alpha}+t_{j,\alpha}^\dagger s_j\right) \\
    &- {\sum_{\langle i j \rangle}} \epsilon_{\alpha\beta\gamma}t_{i,\beta}^\dagger t_{i,\gamma}\epsilon_{\alpha\beta{'}\gamma{'}}t_{j,\beta{'}}^\dagger t_{j,\gamma{'}},
\end{split}
\end{equation}
\begin{align}
        \mathcal{H}_J={\sum_i}\left(-3 s_i^\dagger s_i+\sum_{\alpha}t_{i,\alpha}^\dagger t_{i,\alpha}\right).
\end{align}

In the dimerized phase, the $s_i$ boson condenses. We can replace the operator $s_i$ and $s_i^\dagger$ by a real number $\bar{s}$. Furthermore, we replace the local constraint on each {vertical} bond by a global one $\sum_i s_i^\dagger s_i+\sum_{i, \alpha} t_{i,\alpha}^\dagger t_{i, \alpha} = 
{\Nd}
$ with
{$\Nd$ being the number of dimers,}
from which we obtain
$\bar{s}{\simeq}1-\frac{1}{2
  {\Nd}
}\sum_{i, \alpha} t_{i,\alpha}^\dagger t_{i, \alpha}$ {under the Holstein-Primakoff expansion \cite{PhysRevB.69.054423}}. {Here} the collective excitation in the dimerized phase is the triplet excitation, which can be obtained by expanding $\mathcal{H}$ to the 
{quadratic} order in $t_{i, \alpha}$. The dispersion {of the triplet excitation} is given by
\begin{equation}
  {\xi _{\alpha ,k}}
  = \sqrt {2{J_{xy}}{J}{A_k}
    +
    {J}{^2}
  },
\end{equation}
{for $\alpha=x,y$ and}
\begin{equation}
  {\xi _{z ,k}} 
  = \sqrt {2{J_z}{J}{A_k} +
    {J}{^2}
  }, 
\end{equation}
where $A_k=\cos k_x+\cos k_y$. {The higher order terms that are responsible for the damping of the Higgs mode are not considered here, but will be included effectively in the field theoretical treatment below.}  The spin anisotropy splits the otherwise triply degenerate triplet excitations into two modes with one having double degeneracy. The gap of $t_z$ triplet excitation first vanishes at
{$(J_z/J)_c = 1/4$} {and $\bm{G}_0=(\pi,\pi)$}
indicating a phase transition into
{the antiferromagnetic}
magnetically ordered phase with spins pointing in the
{$\pm$}
$z$ direction.

\begin{figure}[t]
  \begin{center}
    \includegraphics[width=\columnwidth,bb=0 0 730 952]{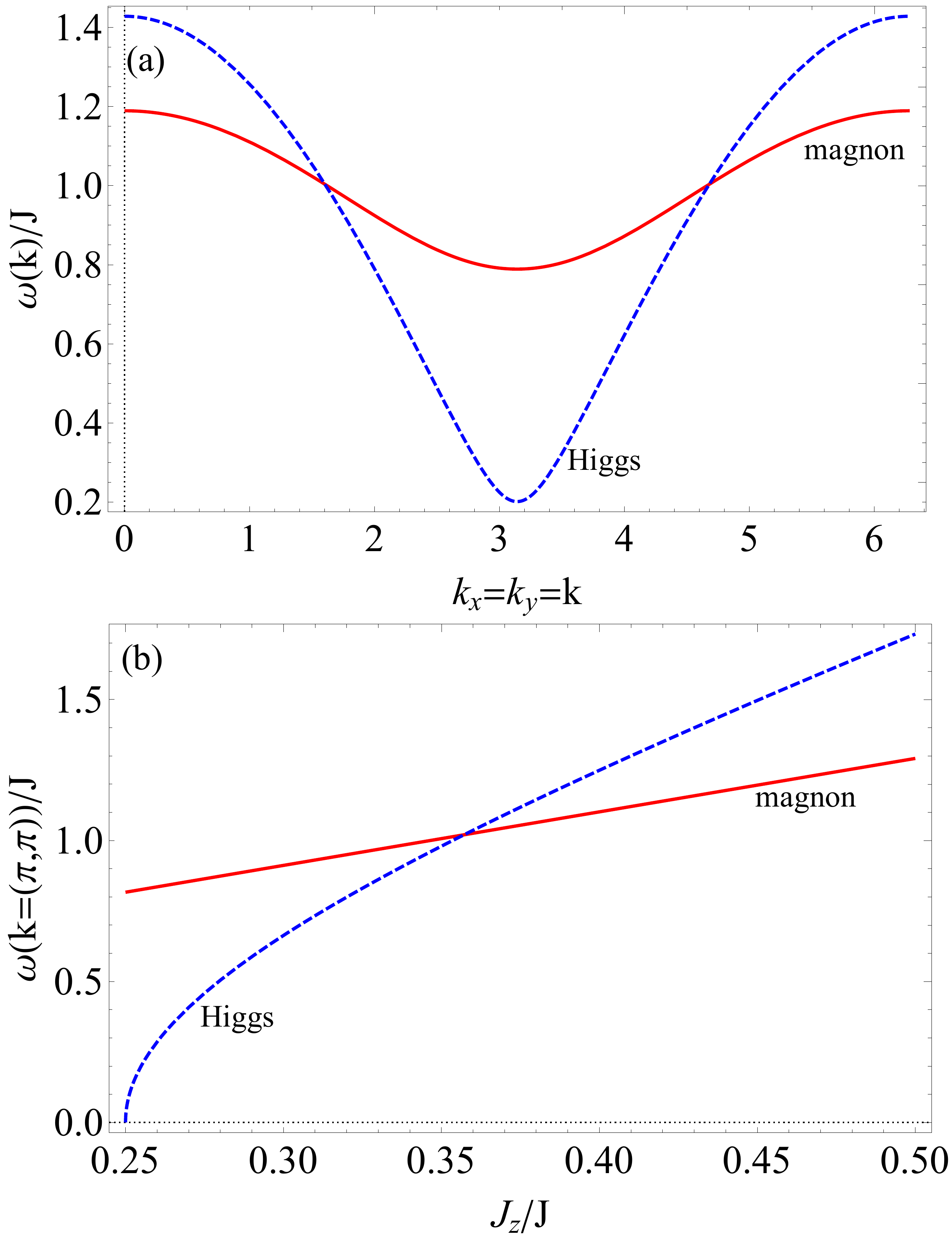}
  \end{center}
\caption{(a) Dispersion of the Higgs and magnon mode for $J_{xy}=0.1{J}$ and $J_z=0.255{J}$, and (b) the corresponding gap at the wavevector $\mathbf{G}_0$ for $J_z/J_{xy}=3$.} 
  \label{f2}
\end{figure}

The phase transition point can also be determined by considering the magnetically ordered phase. In this phase, both $s_i$ and $t_{z,i}$
{bosons}
condense. Because of the intralayer AFM interaction, the ordering wave {vector} for the $t_{z,i}$ boson is $\mathbf{G}_0=(\pi,\ \pi)$. The ground state wave function can be approximated by $\ket{\phi_{\mathrm{AFM}}}=\prod_i\ket{\phi_i}$ with 
\begin{align}\label{eq13}
    \ket{\phi_i}=\frac{1}{\sqrt {1 + {\lambda ^2}}}\left( {s_i^\dag  + \lambda \exp \left( {{i}{\mathbf{G}_0}\cdot{\mathbf{r}_i}} \right)t_{i,z}^\dag } \right)\ket{0},
\end{align}
where $\lambda$ is a variational parameter to be determined later. \cite{sommer_magnetic_2001} We can introduce a new basis $\tilde{s}_i^{\dagger}\ket{0}=\ket{\phi _{i}}$. Then the ground state corresponds to the condensation of $\tilde{s}_i^{\dagger}$ boson. The other three operators in this new basis are given by $\tilde{t}_{i, x/y}^{{\dagger}}={t}_{i, x/y}^{{\dagger}}$ and 
\begin{align}
  \tilde{t}_{i, z}^{{\dagger}}=\frac{1}{\sqrt {1 + {\lambda ^2}}}\left( { {-} \lambda \exp \left( {{-i}\mathbf{G}_0\cdot{\mathbf{r}_i}} \right)s_i^\dag  +t_{i,z}^\dag } \right).  
\end{align}
Here $\lambda$ can be determined by minimizing the Hamiltonian. Again we approximate the local constraint for $\tilde{s}_i$ and $ \tilde{t}_{i, \alpha}$ by the global one as in the case of dimerized phase. $\mathcal{H}$ can be expanded to the second order in $\tilde{t}_{i, \alpha}$, $\mathcal{H}=\mathcal{H}_0+\mathcal{H}_1+\mathcal{H}_2$, with 
\begin{align}
  \mathcal{H}_0=\frac{-4
    {\Nd}
    \lambda^2 J_z}{(1+\lambda^2)^2}+\frac{
    {\Nd}
  }{4}\frac{\lambda^2-3}{1+\lambda^2}{J},
\end{align}
\begin{align}
  \mathcal{H}_1=\left[\frac{
      {J}
    }{1+\lambda^2}-\frac{4J_z(1-\lambda^2)}{(1+\lambda^2)^2}\right]\lambda\sum_{i}\exp \left( {{i}\mathbf{G}_0\cdot{\mathbf{r}_i}} \right)\left(\tilde{t}_{i, z}+\tilde{t}_{i, z}^\dagger\right).
\end{align}
The ground state condition requires that the terms linear in $ \tilde{t}_{i, \alpha}$ vanishes, which yields
$\lambda=\sqrt{(4J_z-
  {J}
  )/(4J_z+
  {J}
  )}$.
This $\lambda$ also minimizes $\mathcal{H}_0$ simultaneously. 

The phase diagram can be obtained from the staggered magnetization
\begin{align}
  M_\alpha(G_0)
  &=
  {\frac{1}{\Nd}}
  \bra{\phi_{\mathrm{AFM}}}\sum_i(S_{1,i}^\alpha-S_{2,i}^\alpha)\exp(i \mathbf{G}_0\cdot{\mathbf{r}_i})\ket{\phi_{\mathrm{AFM}}}.
\end{align}
Here $M_x=M_y=0$ and $M_z(G_0)={\sqrt{16 J_z^2-
    {J^2}
}}/{4J_z}$
vanishes continuously at $J_z/J=1/4$ upon decreasing $J_z$, therefore the quantum phase transition is of second order.
{This mean-field}
critical point is independent
{of}
$J_{xy}$ {as long as} $J_{xy}\le J_z$. The phase transition point is consistent with the previous estimate based on the triplet excitation gap.  This consistency is achieved by the approximation scheme used here. First we replace the local constraint by the global one, known as the Holstein-Primakoff approximation (HPA) \cite{PhysRevB.69.054423}. Within HPA, $\langle s_i\rangle=\langle \tilde{s}_i\rangle=1$, therefore the HPA neglects the suppression of the amplitude of the singlet condensate due to the triplet quantum fluctuations. Secondly, we have introduced a rotated basis to describe the magnetically ordered phase. \cite{sommer_magnetic_2001} At $J_z/J=1/4$, the ground state described by Eq. \eqref{eq13} is the same as the dimerized phase. Therefore, the rotated basis connects
{continuously}
to the un-rotated one upon varying $J_z$.
Alternatively, one can introduce a chemical potential $\mu$ to impose the local constraint by adding a term $-\mu (s_i^\dagger s_i+\sum_{\alpha={x, y, z}}t_{i, \alpha}^\dagger t_{i, \alpha}-1)$ to the Hamiltonian {\cite{PhysRevB.49.8901}}. In the magnetically ordered phase, one can assume the condensation of $s_i$ and $t_\alpha$ bosons without introducing the rotated basis. This approximation, however, does not yield the same transition point by treating the dimerized and magnetic phase separately.

The second order contribution $\mathcal{H}_2$ is 
\begin{align}
    \mathcal{H}_2 =\frac{{J^2}}{32J_z}\sum_{\langle i j \rangle}\left(\tilde{t}_{i,z}+\tilde{t}_{i,z}^\dagger\right)\left(\tilde{t}_{j,z}+\tilde{t}_{j,z}^\dagger\right)\nonumber\\
    +\frac{J_{xy}}{2}\sum_{\langle i j \rangle; \alpha=x, y; {\eta}=\pm}\left(\frac{1}{2}+{\eta}\frac{{J}}{8J_z}\right)\left(\tilde{t}_{i,\alpha}+\eta\tilde{t}_{i,\alpha}^\dagger\right)\left(\tilde{t}_{j,\alpha}+\eta\tilde{t}_{j,\alpha}^\dagger\right)\nonumber\\
    +\left(2J_z-\frac{{J}}{2}\right)\sum_i \tilde{t}_{i,z}^\dagger\tilde{t}_{i,z}+\left(2J_z+\frac{{J}}{2}\right)\sum_{i;\alpha=x, y, z} \tilde{t}_{i,\alpha}^\dagger\tilde{t}_{i,\alpha}.
\end{align}
The magnon dispersion associated with the operator $\tilde{t}_{i, x/y}^{\dagger}$ in the AFM phase can be obtained by the Bogoliubov transformation and is
\begin{equation}\label{wm}
    \omega_M=\sqrt{\left(\frac{{J}}{2}+2J_z+\frac{A_k {J} J_{xy}}{4J_z}\right)^2-(J_{xy}A_k)^2}.
\end{equation}
The Higgs mode corresponds to the excitation of $\tilde{t}_{i, z}$ boson and its dispersion is given by
\begin{equation}\label{wh}
    \omega_H=\sqrt{4J_z\left(4J_z+\frac{{J^2}A_k}{8J_z}\right)}.
\end{equation}
The gap of the Higgs mode vanishes at the transition point.
As shown in Fig. \ref{f2}, for a strong anisotropy $\gamma=J_z/J_{xy}$, the Higgs mode can lie below the magnon continuum. In this case, the decay of the Higgs mode into
{magnon}
continuum is expected to be suppressed and therefore the Higgs mode is stabilized. The magnon and Higgs modes cease to exist when the system is tuned to the dimerized phase.

The gaps of the Higgs and magnon modes can be estimated in the Ising limit, $J_z/J_{xy}\rightarrow\infty$. The magnon carries quantum spin number $S_m=\pm 1$, and it corresponds to a single spin flip. The energy cost is $E_m=2J_z+J/2$. The Higgs mode has quantum spin number $S_h=0$, and therefore it corresponds to flip a pair of antiferromagnetically aligned spins between different layers. Its energy cost is $E_h=4J_z$. This simple estimate of the magnon and Higgs gaps agrees well with the results in  Fig. \ref{f2} (b) when the system is in the well developed magnetically ordered phase (large $J_z/J$ region).

We proceed to calculate the dynamic spin structure factor that can be accessed experimentally,
\begin{align}
    \chi_{\alpha}(\omega, \mathbf{q})=\sum_{l=1,2}\langle S_l^\alpha(-\omega, -\mathbf{q})S_l^\alpha(\omega, \mathbf{q})\rangle_Q,
\end{align}
where $\langle\dots \rangle_Q$ denotes quantum average. Knowing the dispersion for the magnon and Higgs modes, $\chi_{\alpha}$ can be obtained straightforwardly
\begin{align}
    \chi_{\alpha}=C_k \left(\frac{1}{\omega+i0^+ + \omega_M }-\frac{1}{\omega+i0^+- \omega_M}\right),\\
    C_k=\frac{1}{64\omega_M}\left(4+\frac{{J}}{J_z}\right)\left[2{J}+8J_z+A_k J_{xy}\left(\frac{{J}}{J_z}-4\right)\right],
\end{align}
for $\alpha=x, y$ and $0^+$ represents a positive infinitesimal number. For $\chi_z$, we have
\begin{align}
   \chi_{z}= \frac{{J^2}}{8 J_z \omega_H} \left(\frac{1}{\omega+i0^+ + \omega_H }-\frac{1}{\omega+i0^+- \omega_H}\right).
\end{align}
The magnon (Higgs) excitation appears in the transverse (longitudinal) susceptibility. No damping has been taken into account here so the spectral density is a delta function.  

The Higgs peak in $\chi_\alpha$ can be smeared out severely in the presence of decay, especially in low dimensional systems. To detect the Higgs mode, singlet bond susceptibility was introduced and was shown to exhibit a sharp Higgs peak despite of the strong damping \cite{PhysRevB.92.245137}. The singlet bond susceptibility is analogous to the scalar susceptibility introduced in Ref. \onlinecite{PodolskyArovasPRB}.  The singlet bond susceptibility is defined as
\begin{align}
    \chi_B(\omega, \mathbf{q})=\langle B(-\omega, -\mathbf{q})B(\omega, \mathbf{q})\rangle_Q,
\end{align}
with $B_i=\mathbf{S}_{1,i}\cdot \mathbf{S}_{2,i}$. It can be calculated
\begin{align}
  \chi_B=  \frac{({J}+2J_z)^2}{64J_z^2}\delta(\mathbf{q})\delta(\omega)+(16J_z^2-{J^2})\chi_z(\omega, \mathbf{q}-\mathbf{G}_0).
\end{align}
The first term accounts for the static dimer correlation at $Q=0$. There is a $\mathbf{G}_0$ momentum shift between $\chi_B$ and $\chi_z$ because of the N\'{e}el order. Approaching the quantum critical point, the spectral density of $\chi_B$ vanishes as $\sqrt{16 J_z^2-{J^2}}$.

\begin{figure}[t]
  \begin{center}
    \includegraphics[width=\columnwidth,bb=0 0 393 162]{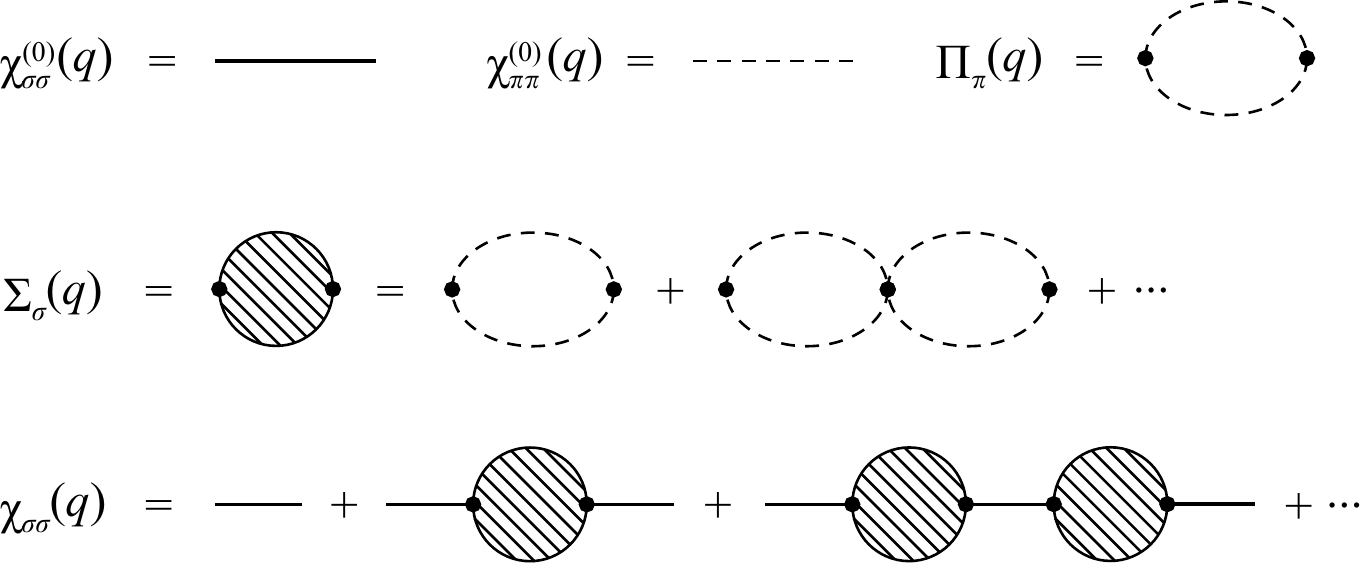}
  \end{center}
\caption{Feynman diagrams describing the self energy $\Sigma_\sigma(q)$ under the random phase approximation of the polarization bubble $\Pi_\pi(q)$ and the full susceptibility of Higgs mode $\chi_{\sigma\sigma}(q)$. The solid and dashed lines represent the bare susceptibility of Higgs and magnon modes, respectively.} 
  \label{f3}
\end{figure}

\begin{figure*}[t]
  \begin{center}
    \includegraphics[width=15cm]{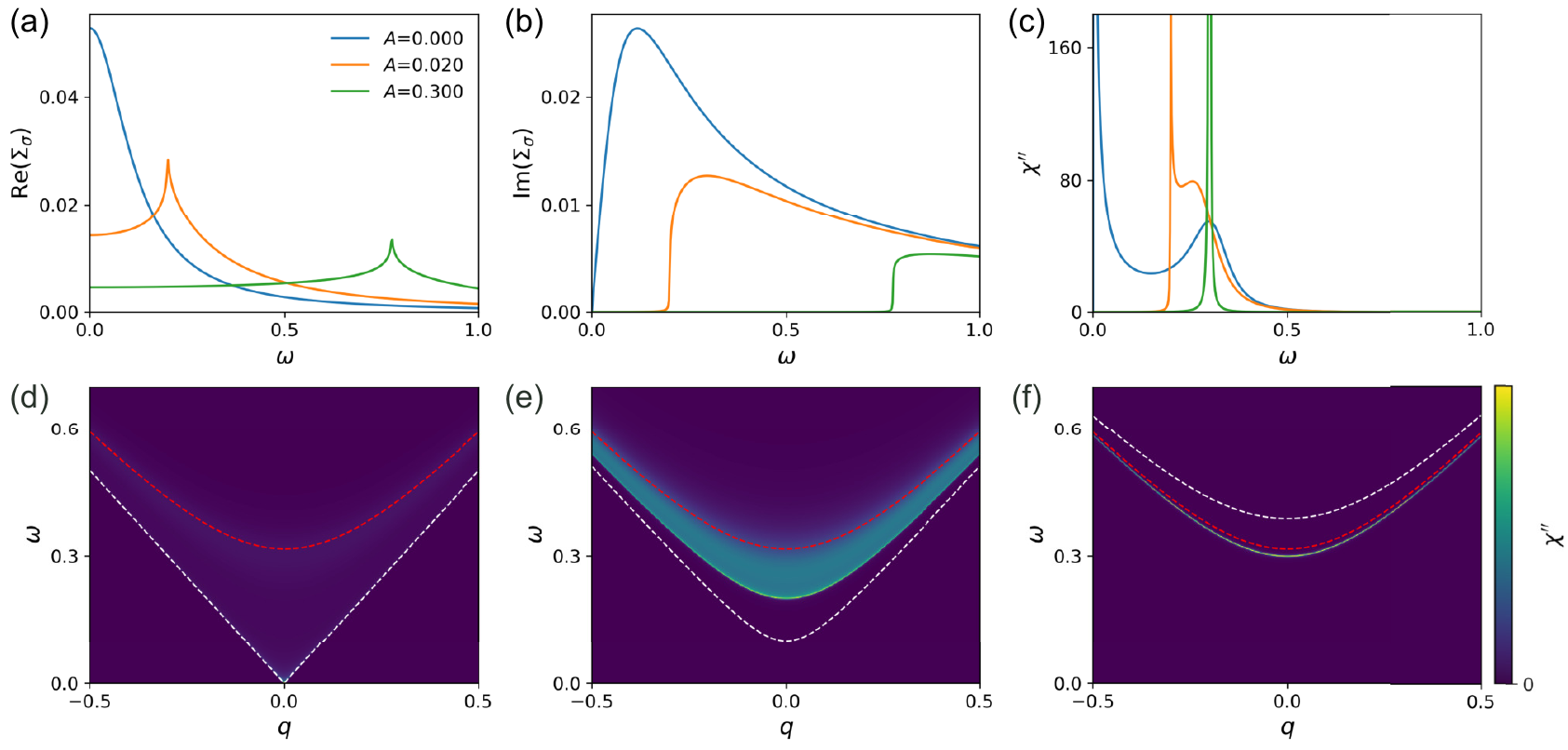}
  \end{center}
  \caption{(a) and (b) Real and imaginary parts of the self energy of Higgs mode {at $\bm{q}=\bm{0}$ and} for $D=2+1$,
    $N=100$, $\Lambda=3m_0$, and $g=0.9g_c$. Here we consider three different anisotropies with $A=0,$ 0.02, and 0.3
            {[see Eq.~\eqref{S}]}. (c) The corresponding spectral function at $\bm{q}=\bm{0}$. (d)-(f) The intensity plot of the spectral functions of the Higgs mode for $A=0,$ 0.02, and 0.3, respectively. Here the  red and white dash lines are the bare dispersion of Higgs and magnon modes. {The decay of the Higgs mode to the magnon modes smears the spectral peak of the Higgs mode in (d) and (e). Because the magnon mode is above the Higgs mode in (f), the spectral peak of the Higgs mode is sharp}.} 
  \label{f4}
\end{figure*}

\section{Field theoretical approach}
In the mean-field bond operator approach, we have shown that the magnon modes can be gapped due to the magnetic anisotropy and the magnon
{energy}
can {be even larger than that of} the
{long-wavelength Higgs mode.}
The question is how the magnon gap affects the lifetime of the Higgs mode. Here we proceed to calculate the lifetime of the Higgs mode by considering the decay of the Higgs mode into magnon modes. A more convenient method is
{a}
field theoretical approach based on an effective action. We generalize the calculations in Ref. \onlinecite{PodolskyArovasPRB} by including the spin anisotropy. We consider
{an action of}
the relativistic $\mathcal{O}(N)$ field theory with anisotropy, which describes various condensed matter systems.
For example{, the case with $N = 3$}
describes the long wavelength fluctuation in the {anisotropic}
Heisenberg model. The Euclidean time action of the model reads as    
\begin{align}\label{S}
  \mathcal{S}=\frac{1}{2g}\int_\Lambda d^D x \left[ (\partial_{\alpha}\bm{\Phi})^2+\frac{m_0^2}{4N}\left(|\bm{\Phi}|^2-N\right)^2+\frac{A}{2}\sum_{i=2}^{N}\Phi_i^2\right],
\end{align}
where $\bm{\Phi}$ is a $N$-component vector {field}, which can be parametrized by $\bm{\Phi}=(\Phi_1, \bm{\pi})$ with $\bm{\pi}$ being the $(N-1)$-component vector. $D=d+1$ is the space-time dimension. $A>0$ is the hard axis anisotropy, which ensures the saddle point solution 
{$\langle\Phi_1\rangle=\sqrt{N}$} and 
{$\langle\bm{\pi}\rangle=\bm{0}$}. We do not write the anisotropy in the easy axis anisotropy form $-A \Phi_1^2 /2$ because the saddle point solution depends on $A$ in this case. Here $m_0$ is the bare mass and $\Lambda$ is the ultraviolet cutoff wavevector, both of which depend on the microscopic details of the systems. $g$ is a  parameter which controls the strength of quantum fluctuations. There exists a quantum phase transition at $g=g_c$ and the system orders when $g<g_c$. Because of the anisotropy $A>0$, the phase transition exists at $d{\geq}1$.

The fluctuations {of the field} in the ordered phase can be parametrized as
\begin{align}
   \bm{\Phi}=(r\sqrt{N}+\sigma, \bm{\pi}),
\end{align}
where $r$ is responsible for the suppression of the order parameter due to the quantum fluctuation. The action Eq.~\eqref{S} can be expanded as
\begin{align}
    \mathcal{S}=\mathcal{S}_0+\mathcal{S}_A+\mathcal{S}_C,\\
    \mathcal{S}_0=\frac{1}{2g}\int _{\Lambda }d^Dx\left[\left(\partial_\mu \sigma \right)^2+\left(\partial_\mu \bm{\pi} \right)^2+m_0^2 r^2 \sigma ^2+\frac{ A}{2}\pi ^2\right],\\
    \mathcal{S}_A=\frac{m_0^2}{2 g}\int _{\Lambda }d^Dx\left[\frac{r \left(\sigma ^3+\sigma \bm{\pi} ^2\right)}{\sqrt{N}}+\frac{\left(\sigma ^2+\bm{\pi} ^2\right)^2}{4 N}\right],\\
    \mathcal{S}_C=\frac{m_0^2 \left(r^2-1\right)}{4 g}\int _{\Lambda }d^Dx\left(2 \sqrt{N} r \sigma +\sigma ^2+\bm{\pi} ^2\right),
\end{align}
where {$\mathcal{S}_0$ is the free field action with anisotropy}, $\mathcal{S}_A$ {collects the anharmonic contributions,} and $\mathcal{S}_C$ is the counterterm. 
{The bare susceptibility of Higgs and magnon modes from $\mathcal{S}_0$ are
\begin{align}\label{bs}
    \chi _{\sigma\sigma}^{(0)}(q)=\frac{g}{q^2+ r^2 m_0^2}, \qquad \chi _{\pi\pi}^{(0)}(q)=\frac{g}{q^2+{A}/{2}},
\end{align}
where $rm_0$ is the renormalized mass of the Higgs mode
{and $q$ denotes a $D$-dimensional momentum}. Under the analytical continuation $q^2\rightarrow \mathbf{q}^2-(\omega+i 0^+)^2$, the zeroth order Higgs and magnon dispersions are
$\omega_H^{(0)}
{(\mathbf{q})}
=\sqrt{\mathbf{q}^2+m_0^2r^2}$ and
$\omega_M^{(0)}
{(\mathbf{q})}
=\sqrt{\mathbf{q}^2+A/2}$, respectively. {The dispersions can be viewed as the expansion of Eq. (\ref{wm}) and (\ref{wh}) around $\mathbf{G}_0$ up to a renormalization factor. Then we can make the correspondence that $A\simeq 2J/J_{xy}-8$ and $rm_0 \simeq \sqrt{J/J_z-4}$ around the mean-field QCP $(J_z/J)_c=1/4$ .} The Higgs gap, $rm_0$, vanishes at the QCP at  $g=g_c$, consistent with the previous bond operator approach.  Because the magnon gap remains {nonzero with $\omega_M^{(0)}(0)=$}
$\sqrt{A/2}$, the Higgs mode is below the magnon mode in the long wavelength limit near the QCP considered here. 
}

To ensure $\bm{\Phi}_g=(r\sqrt{N},\bm{0})$ is a stable ground state, the expectation of $\sigma$ must vanish, $\langle\sigma\rangle=0$. This means that the sum of all one-particle irreducible (1PI) diagrams with one $\sigma$ {external} leg mush vanish. In the large $N\gg 1$ limit, {the cancellation of the two leading-order 1PI diagrams originated from the term $r\sigma\bm{\pi}^2/\sqrt{N}$ in $\mathcal{S}_A$ and $2 \sqrt{N} r \sigma $ in $\mathcal{S}_c$ yields} $r=\sqrt{1-g/g_c}$ with
\begin{equation}\label{gc}
\begin{split}
g_c&=\left[\int_\Lambda \frac{d^D k}{(2\pi)^D}\frac{1}{k^2+A/2}\right]^{-1} \\
&=\begin{cases}
\frac{4\pi}{\sqrt{\Lambda^2+A/2}-\sqrt{A/2}}, & D=3  \\
\\
\frac{8\pi^2}{\Lambda\sqrt{\Lambda^2+A/2} - \frac{A}{2} \ln\frac{\Lambda+\sqrt{\Lambda^2+A/2}}{\sqrt{A/2}}}, & D=4
\end{cases}
\end{split}
\end{equation}
where the integral is due to the $\pi$ loop contribution in the term $r\sigma\pi^2/\sqrt{N}$ {(see Appendix \ref{AA})}. In the limit $\Lambda\gg A$, $g_c=4\pi/\Lambda$ for $D=3$ and $g_c=8\pi^2/\Lambda^2$ for $D=4$. Here $g_c$ depends only on the ultraviolet cutoff {but not} the easy axis anisotropy.

The full Higgs mode susceptibility is given by the Dyson equation 
\begin{align}
    \chi _{\sigma \sigma }(q)=\frac{g}{q^2+m_0^2 r^2-g \Sigma _{\sigma }(q)},
\end{align}
where $\Sigma _{\sigma }(q)$ is the self-energy that collects all the 1PI diagrams. In the one-loop order, we consider the {dominant} polarization bubble {from the $r\sigma\bm{\pi}^2/\sqrt{N}$ term in $\mathcal{S}_A$ under the second order perturbation expansion} 
\begin{widetext}
\begin{align}\label{Pi}
    \Pi _{\pi } (q)=\frac{m_0^4  r^2}{2 }\int _{\Lambda }\frac{d^Dk}{(2 \pi )^D}\frac{1}{\left(k^2+A/2\right)\left[(k+q)^2+A/2\right]} = \frac{m_0^4 r^2}{2}\begin{cases}
               \frac{1}{4\pi\sqrt{q^2}}\cot^{-1}\left(\sqrt{\frac{2A}{q^2} }\right), & D=3\\
               \frac{1}{16\pi^2}\left[1+\log\left(\frac{2\Lambda^2}{A}\right)  -2\sqrt{\frac{2A+q^2}{q^2}}\tanh^{-1}\sqrt{\frac{q^2}{2A+q^2} }\right], & D=4
            \end{cases}
\end{align}
\end{widetext}
that describes the decay of one Higgs mode into two magnon modes, {as shown in Fig. \ref{f3}}. {The loop integral can be evaluated by using the Feynman parameterization (see Appendix \ref{BB}).} The decay of one Higgs mode into other Higgs modes is negligible in the large $N$ limit {and has no contribution in the low frequency region} \cite{PodolskyArovasPRB}. The {other} one-loop tadpole diagrams {from $\sigma^4$ and $2\sigma^2\bm{\pi}^2$ in $\mathcal{S}_A$} are cancelled out by {$(r^2-1)\sigma^2$ in} the counterterm {$\mathcal{S}_C$, as shown in Appendix \ref{CC}}.
Going beyond the one-loop order, we introduce the random phase approximation (RPA) of bubble diagrams and the self energy becomes
\begin{align}
    \Sigma _{\sigma }(q)=\frac{\Pi _{\pi } (q)}{1+ g\Pi _{\pi } (q)/m_0^2 r^2},
\end{align}
{as shown in Fig. \ref{f3}}. The spectral function {of the Higgs mode} is 
\begin{equation}
\begin{split}
\chi''_{\sigma\sigma}(q) & \equiv \mathrm{Im}[\chi_{\sigma\sigma}(q)] \\
& =  \frac{g^2\text{Im}[\Sigma_\sigma(q)]}{\left(q^2+r^2m_0^2-g\text{Re}[\Sigma_\sigma(q)]\right)^2 + g^2\text{Im}[\Sigma_\sigma(q)]^2},
\end{split}
\end{equation}
which is a Lorentzian function. {For $\textbf{q}=\bm{0}$}, the spectral peak is centered at
$\omega_c=\sqrt{r^2m_0^2-g \mathrm{Re}[\Sigma_\sigma(\omega_c)]}$ and its width is $\Gamma_\sigma=2g\mathrm{Im}[\Sigma_\sigma(\omega_c)]$. When twice of the magnon gap is above the Higgs mode {for $A>2r^2m_0^2$}, the decay of the Higgs mode into the magnon mode is absent. Since the Higgs mode becomes the lowest lying mode in this case, there is no decay channel of the Higgs mode even including the higher order processes. Therefore, the Higgs mode can be stable in anisotropic quantum magnets. 

\begin{figure*}[t]
  \begin{center}
    \includegraphics[width=15cm]{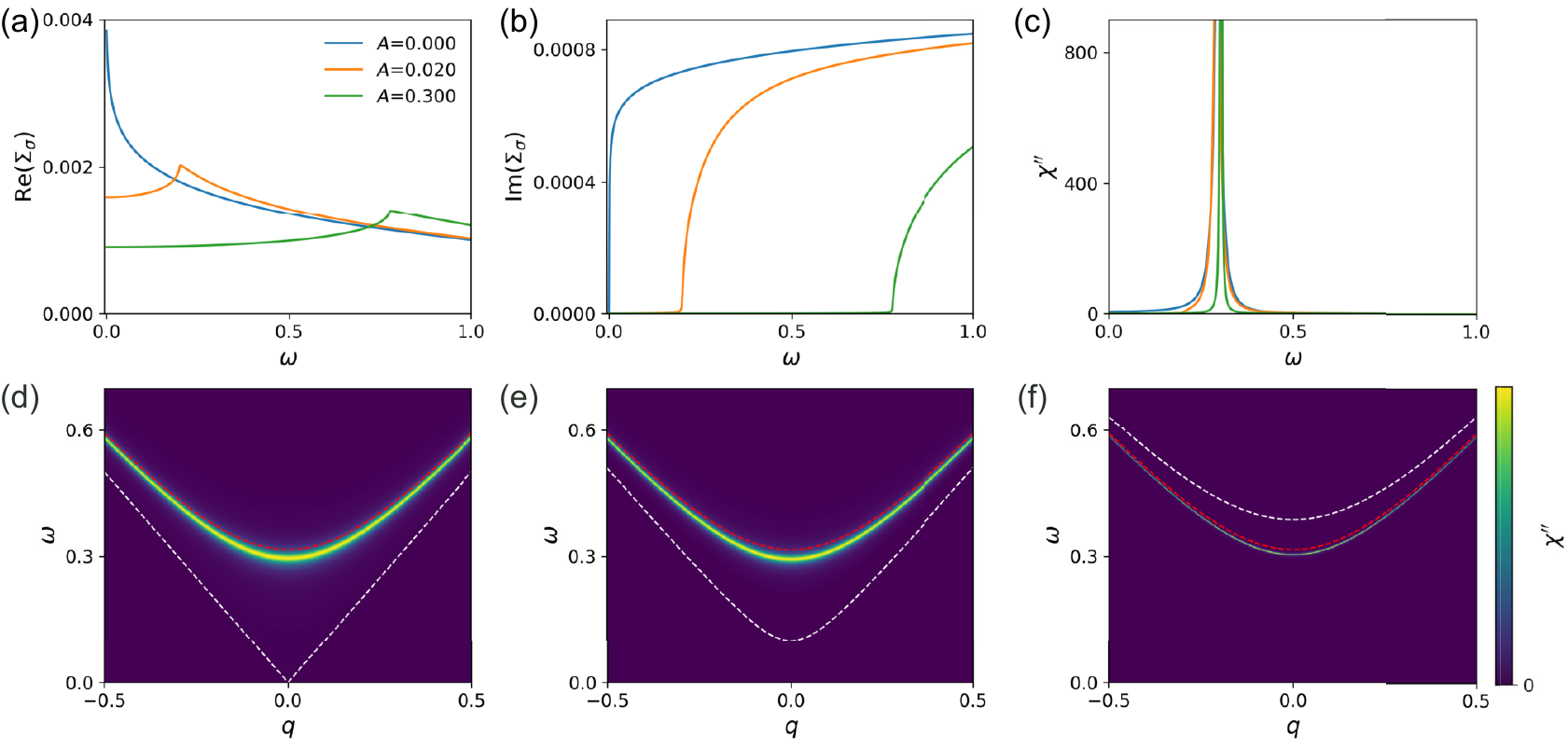}
  \end{center}
\caption{(a) and (b) Real and imaginary parts of the self energy of Higgs mode {at $\bm{q}=\bm{0}$ and} $D=3+1$. The other parameters are same as those used in Fig. \ref{f4}.  (c) The spectral function at $\bm{q}=\bm{0}$. (d)-(f) The intensity plot of the spectral functions of the Higgs mode for $A=0,$ 0.02, and 0.3, respectively, with the red and white dash lines representing the bare dispersion of Higgs and magnon modes. {The spectral peak of the Higgs mode is sharp in $D=3+1$.}} 
  \label{f5}
\end{figure*}

In Figs. \ref{f4}, we show the self energy and spectral function of the Higgs mode for $D=3$. At $\bm{q}=\bm{0}$, the finite ${\rm Re}(\Sigma_\sigma)$ in Fig. \ref{f4}(a) shifts the spectral peak downward slightly [see Fig. \ref{f4}(f)]. For $A<2r^2m_0^2$, the ${\rm Im}(\Sigma_\sigma)$ shown in Fig. \ref{f4}(b) remains finite at $\omega_c$, which broadens the spectral peak due to the decay of Higgs mode into magnon modes, as shown in Figs. \ref{f4}(c)-\ref{f4}(e). On the other hand, when  $A>2r^2m_0^2$,  ${\rm Im}(\Sigma_\sigma)$ vanishes at $\omega_c$ due to the absence of a decay channel. In this case, a pronounced spectral peak of Higgs mode is identified as displayed in Figs. \ref{f4}(c) and \ref{f4}(f). Here we add a tiny imaginary part to the frequency such that the spectral peak for $A>2r^2m_0^2$ has a finite width. The self energy and spectral function for $D=4$, shown in Fig. \ref{f5}, are similar to that for $D=3$. However, due to the weaker quantum fluctuation in higher dimension, the self energy in Figs. \ref{f5}(a) and \ref{f5}(b) is much smaller than that in Figs. \ref{f4}(a) and \ref{f4}(b). As a consequence, even when $A<2r^2m_0^2$, the spectral peak of Higgs mode is still apparent, as shown in Figs. \ref{f5}(c)-\ref{f5}(f). The spectral peak width decreases as $A$ increases to $A=2r^2m_0^2$ above which ${\rm Im}[\Sigma_\sigma(w_c)]=0$ and the spectral peak has zero width.

\section{Quantum Monte Carlo Results}
In the field theoretical calculations, we have considered the large $N\gg 1$ limit in order to make controlled approximation. To connect to physical spin with $N=3$, 
  below, we show the results of our unbiased quantum Monte Carlo simulation to address the excitation spectrum near the QCP in the Ising-like bilayer \textit{XXZ} model.
  Especially, we focus on $d = 2$, where the anisotropy effect to stabilize the Higgs mode is expected as more drastic than $d = 3$.
  Our quantum Monte Carlo simulation is based on the directed-loop algorithm~\cite{PhysRevE.66.046701,PhysRevE.71.036706} with the continuous imaginary time world-line scheme. The analytical continuation from the imaginary to real frequency is the core part of our numerical study, which we perform utilizing the recently developed stochastic optimization method~\cite{PhysRevB.95.014102}.

\begin{figure*}[t]
  \begin{center}
    \includegraphics[width=14cm,bb=0 0 700 481]{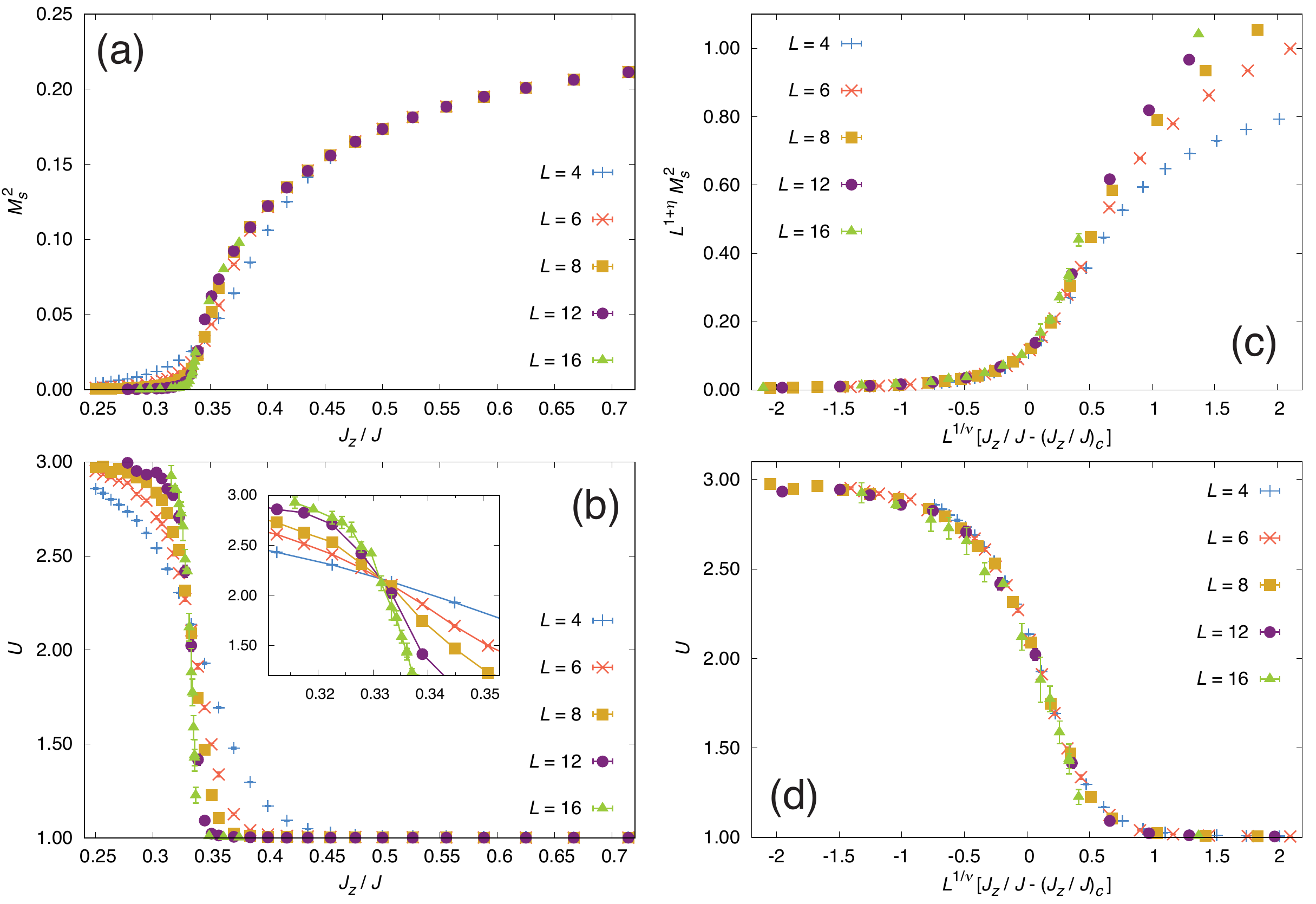}
  \end{center}
  \caption{%
    {
      (a) Quantum Monte Carlo results of $\langle{M_s^2}\rangle$ [Eq.~\eqref{eq:Ms}] and (b) $U$ [Eq.~\eqref{eq:U}] for $J_{z} / J_{xy} = 3$ as a function of $J_z/J$ at $\beta J_z = 2.5 \times L$.
      Finite-size scaling of (c) $\langle{M_s^2}\rangle$ and (d) $U$ where we assume the critical exponents of the $D=2+1$ Ising universality class:
      $\nu = 0.63012(16)$, $\eta = 0.03639(15)$, and $z = 1$~\cite{PhysRevE.65.066127}. $(J_z /J)_c = 0.332(1)$ is obtained by the data collapse of the presented data.
    }
  } 
  \label{f6}
\end{figure*}

We simulated the bilayer square-lattice Hamiltonian [Eq.~\eqref{eq:H}]
by adopting periodic boundary conditions in the $a$ and $b$ directions.
  First, to determine the QCP induced by changing $J_z / J$, we consider a source term of a longitudinal staggered field $-h_s\sum_{l,i}(-1)^{l}e^{i\mathbf{G}_0\cdot r_{i}}S^z_{l,i}$ with $\mathbf{G}_0 = (\pi,\pi)$, thereby define the Binder parameter,
\begin{align}
  U =
  \frac{\langle{M_s^4}\rangle}{ \langle{M_s^2}\rangle^2},
  \label{eq:U}
\end{align}
with
\begin{align}
  \langle{M_s^2}\rangle
  =
  \frac{T^2}{\Ns^2 Z}
  \frac{\partial^2 Z}{\partial h_{s}^2}\Biggr\rvert_{h_{s} = 0},
  ~~~
  \langle{M_s^4}\rangle
  =
  \frac{T^4}{\Ns^4 Z}
  \frac{\partial^4 Z}{\partial h_{s}^4}\Biggr\rvert_{h_{s} = 0},
  \label{eq:Ms}
\end{align}
where $Z$ is the partition function and $\Ns = 2L^2$ is the total number of sites ($L$ denotes the system size). $U$ is a dimensionless scaling parameter and is expected to be asymptotically size independent at the QCP. In Figs.~\ref{f6}(a) and \ref{f6}(b), we show the $J_z /J$ dependence of $\langle{M_s^2}\rangle$ and $U$, respectively, for $J_{z} / J_{xy} = 3$ and $4 \le L \le 16$. To investigate quantum critical behaviors, the inverse temperature $\beta = 1/T$ is set to $\beta J_z = 2.5 \times L$, anticipating the $D=2+1$ Ising universality class where the dynamical scaling exponent is $z = 1$; this temperature is low enough to study ground state properties. We find that $\langle{M_s^2}\rangle$ increases with increasing $J_{z} / J$. Furthermore, in the region where rapid increase of $\langle{M_s^2}\rangle$ suggests a QCP, $U$ shows a clear tendency towards crossing for different $L$. By using the finite-size scaling analysis, we obtain the estimate of the QCP as $(J_z/J)_c = 0.332(1)$ for $J_{z} / J_{xy} = 3$ based on the data collapse shown in Figs.~\ref{f6}(c) and \ref{f6}(d). The enlarged disordered phase relative to the prediction of the mean-field theory, $(J_z/J)_{c,\text{MF}} = 1/4$, is a normal observation.  

To study the excitation spectrum near the QCP, we measure the imaginary-time dynamical correlation function in the quantum Monte Carlo simulation,
\begin{align}
  C^{zz}_{l,ij}(\tau)
  &= \frac{1}{Z} \mathrm{Tr}\, T_\tau
  \left(
  e^{-\beta \mathcal{H}} S^z_{l,i}(\tau) S^z_{l,j}(0)
  \right),
  \\
  C^{xx}_{l,ij}(\tau)
  &= \frac{1}{Z} \mathrm{Tr}\, T_\tau
  \left(
  e^{-\beta \mathcal{H}} S^x_{l,i}(\tau) S^x_{l,j}(0)
  \right),
\end{align}
where $0 \le \tau \le \beta$, $T_\tau$ denotes the time ordering operator, and $S^\alpha_{l,i}(\tau) = e^{\tau\mathcal{H}} S^\alpha_{l,i} e^{-\tau\mathcal{H}}$ ($\alpha = z,x$). We evaluate the $\tau$ dependence of $C^{\alpha\alpha}_{l,ij}(\tau)$ at equally-spaced discrete sample points, $\tau_m = m\Delta\tau$, $0 \le m < N_\tau$, where
$N_\tau \equiv \beta / \Delta\tau$
is taken as $N_\tau = 200L$ in our study.
By taking the input of $C^{\alpha\alpha}_{l,ij}(\tau)$, the stochastic optimization method~\cite{PhysRevB.95.014102} can numerically execute the analytical continuation and the Fourier transformation to yield the corresponding dynamical spin structure factor,
\begin{align}
  C^{\alpha\alpha}_{l,\mathbf{k}}(\omega)
  =
  \frac{1}{2\pi L^2} \sum_{i,j} \int_{-\infty}^{\infty} dt
  \left\langle{
    S^{\alpha}_{l,i}(t) S^{\alpha}_{l,j}(0) 
  }\right\rangle_{T}
  e^{-i[\mathbf{k}\cdot(\mathbf{r}_i - \mathbf{r}_j) - \omega t]},
  \label{eq:Ckw}
\end{align}
with $\alpha = z,x$, $S^\alpha_{l,i}(t) = e^{i\mathcal{H}t} S^\alpha_{l,i} e^{-i\mathcal{H}t}$, and $\langle{\dots}\rangle_T$ denotes thermal average at temperature $T$. Our simulations for these dynamical correlation functions are carried out at $\beta J_z = 0.833 \times L$, which is still a low enough temperature to address the ground-state spectral function.

\begin{figure*}
  \begin{center}
    \includegraphics[width=14cm,bb=0 0 622 607]{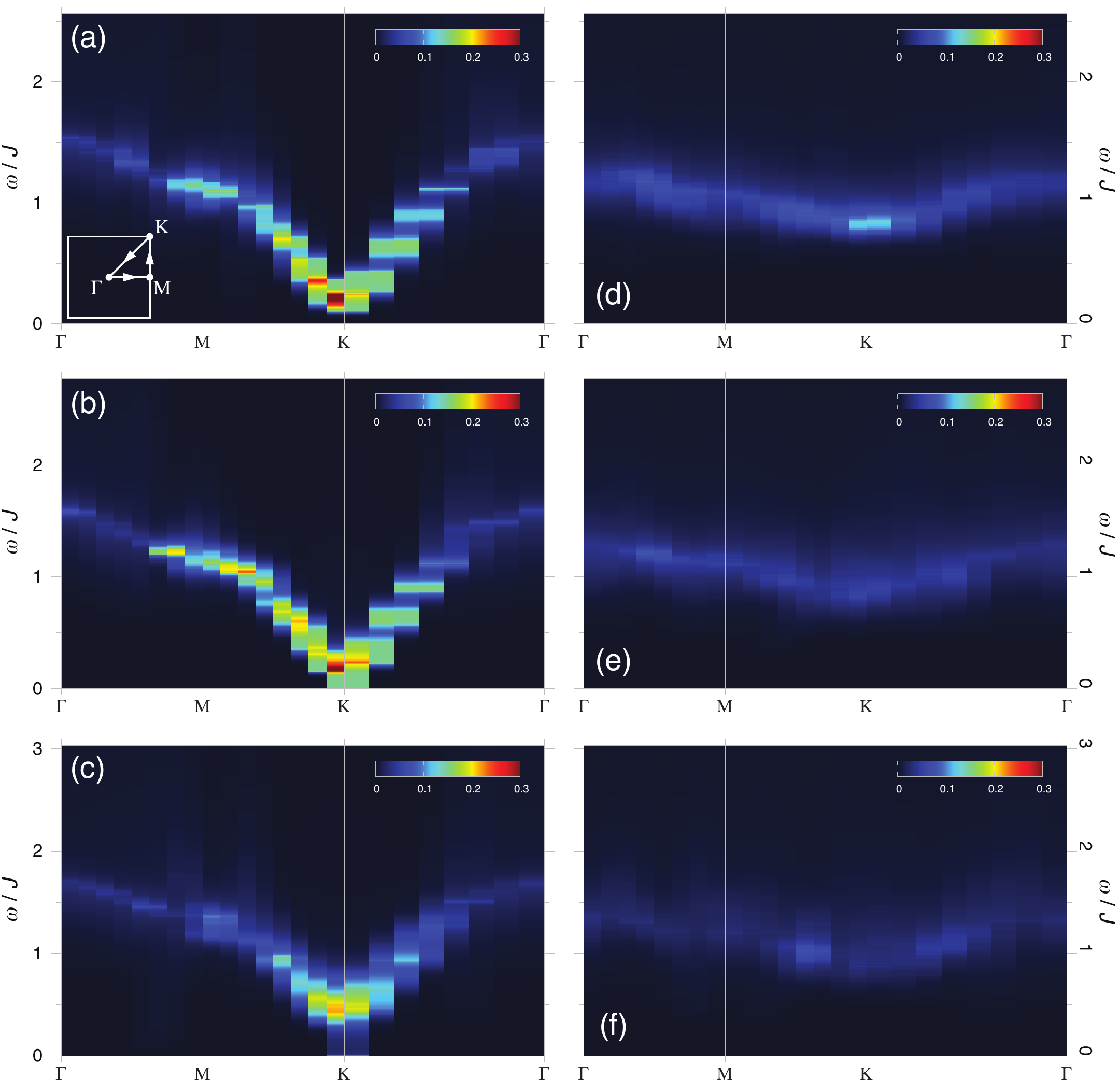}
  \end{center}
  \caption{%
    {%
      Results of $C^{zz}_{l,\mathbf{k}}(\omega)$ [Eq.~\eqref{eq:Ckw}] obtained by the quantum Monte simulation and the analytical continuation
      for $L = 16$ and (a) $J_z / J = 1/3.25 = 0.927(3) \times (J_z/J)_c$, (b) $J_z / J = 1/3 = 1.004(3) \times (J_z/J)_c$, and (c) $J_z / J = 1/2.75 = 1.095(3) \times (J_z/J)_c$, corresponding to the dimerized phase, a vicinity of the QCP, and the magnetically ordered phase, respectively. The results of $C^{xx}_{l,\mathbf{k}}(\omega)$ in the same parameters are shown in (d) $J_z / J = 1/3.25$, (e) $J_z / J = 1/3$, and (f) $J_z / J = 1/2.75$. The results are shown along the line in the Brillouin zone shown in (a).
    }%
  } 
  \label{f7}
\end{figure*}

We show the results of $C^{zz}_{l,\mathbf{k}}(\omega)$ in the intensity plot for $L = 16$ in Figs.~\ref{f7}(a)--\ref{f7}(c). The consistency with the result for $L = 12$ has been checked (not shown). The result in Fig.~\ref{f7}(a) for $J_z / J = 1/3.25 = 0.927(3) \times (J_z/J)_c$ shows the gapped $z$-component of the triplet excitation in the dimerized phase. Figure~\ref{f7}(b) corresponds to the spectrum in the vicinity of the QCP and shows the quantum critical soft mode at $\mathbf{k} = (\pi,\pi)$ for $J_z / J = 1/3 = 1.004(3) \times (J_z/J)_c$. Finally, the result in Fig.~\ref{f7}(c) shows the gapped Higgs excitations in the magnetically ordered phase for $J_z / J = 1/2.75 = 1.095(3) \times (J_z/J)_c$. The observed Higgs excitations are relatively sharp (smeared) in the long-wavelength limit $\mathbf{k} \simeq (\pi,\pi)$ [away from $\mathbf{k} \simeq (\pi,\pi)$], consistent with our field theory predictions.

We also show the results of $C^{xx}_{l,\mathbf{k}}(\omega)$ for the same set of parameters in Figs.~\ref{f7}(d)--\ref{f7}(f). We find that the spectral weight corresponding to the $xy$-components of the triplet excitation in the dimerized phase and the one corresponding to the magnons in the ordered phase seem to evolve continuously into each other by varying $J_z / J$. These excitations are gapped and with small bandwidths all the way through the QCP. Remarkably, by comparing $C^{zz}_{l,\mathbf{k}}(\omega)$ and $C^{{xx}}_{l,\mathbf{k}}(\omega)$ in the ordered phase, we find that the stable Higgs excitations emerge below the gapped magnon band near $\mathbf{k} = (\pi,\pi)$, whereas the smeared Higgs excitations away from $\mathbf{k} \simeq (\pi, \pi)$ are within the energy range of the less dispersive magnon band [Figs.~\ref{f7}(c) and \ref{f7}(f)]. This observation confirms the predicted mechanism of the protection of the long-wavelength Higgs mode through the violation of the kinematic condition near the QCP.

\section{Conclusions}

In this work, we show the existence of a stable Higgs mode in
{an anisotropic quantum spin system near the QCP.}
The easy axis
{anisotropy gaps out}
magnons, while the Higgs mode gap vanishes at the QCP between the
{magnetically ordered}
and dimerized phases. Therefore,
 close to the QCP, the energy of the
{long-wavelength Higgs mode is lower than that of magnons.}
As a consequence, the decay of the Higgs mode to
{{the}} magnon modes is forbidden due to the energy conservation. In the quantum field theory perspective, the  system can be described by a coarse-grained $O(N)$ nonlinear $\sigma$ model \cite{PodolskyArovasPRB} with an easy axis anisotropy term. The anisotropy completely suppresses the damping of Higgs mode above a critical value. In this case, the Higgs mode is the lowest lying mode and becomes stable even in
{$d = 2$. Our quantum Monte Carlo simulation indeed demonstrates the stability of the Higgs mode in the bilayer square-lattice \textit{XXZ} model around $\mathbf{k} = (\pi,\pi)$ near the QCP. Hence, our theory and simulation establish a new} mechanism to stabilize the Higgs mode in anisotropic quantum magnets near a QCP, which can be an ideal platform to study the Higgs physics.

\begin{acknowledgments}
  The authors would like to thank Anders W. Sandvik, Cristian D. Batista, Ziyang Meng, Marc Janoschek, and Filip Ronning for helpful discussions. This work was carried out under the auspices of the U.S. DOE NNSA under contract No. 89233218CNA000001 through the LDRD Program and the U.S. DOE Office of Basic Energy Sciences Program E3B5 (S.-Z. L. and J.-X. Z.). A.M.-K. used computational resources of the HPCI system
    through the HPCI System Research Project (Project IDs.:~hp170213, hp180098, and hp180129).
    Y.K. acknowledges the support from NSFC Research Fund for International Young Scientists No.~11950410507
    as well as the support from Ministry of Science and Technology (MOST) with the Grants No.~2016YFA0300500 and No.~2016YFA0300501.
  W.Z. is supported by the start-up funding from Westlake University.
\end{acknowledgments}


\appendix
\section{Conditions for $\langle\sigma\rangle=0$}\label{AA}
The stable ground state $\bm{\Phi}_g=(r\sqrt{N},\bm{0})$ condition requires
\begin{equation}
\langle \sigma \rangle = \frac{\int \mathcal{D}\bm{\Phi}(x)  \sigma(x)  e^{-\mathcal{S}} }{Z}=0,
\end{equation} 
where $\int \mathcal{D}\bm{\Phi}(x)$ denotes the functional integral and $Z=\int \mathcal{D}\bm{\Phi}(x) e^{-\mathcal{S}}$ is the partition function. Under the perturbation expansion to the leading order and in the large $N$ limit, the sum of the two 1PI diagrams originated from the term $r\sigma\pi^2/\sqrt{N}$ in $\mathcal{S}_A$ and $2 \sqrt{N} r \sigma $ in $\mathcal{S}_c$ must vanish as 
\begin{equation}\label{couter}
\frac{m_0^2 r}{2 g \sqrt{N}} \int d^D x' G_{\sigma\sigma}^{(0)}(x-x')  \left[(N-1)G_{\pi\pi}^{(0)}(0) + (r^2-1)N \right] = 0,
\end{equation}
where $G^{(0)}_{\alpha\alpha}(x) = \langle \alpha(x)\alpha(0)\rangle_0$ is the bare propagator for $\alpha=\sigma$ or $\pi$.  Eq. (\ref{couter}) yields 
\begin{equation}\label{r2}
r^2 = 1 - g \int_\Lambda \frac{d^D k}{(2\pi)^D}\frac{1}{k^2+A/2} = 1 - \frac{g}{g_c},
\end{equation}
that vanishes at the QCP $g=g_c$ and gives Eq. (\ref{gc}). Here we have used the identity $G_{\pi\pi}^0(0)=(2\pi)^{-D} \int d^D k \chi_{\pi\pi}^0(k)$, where $\chi_{\pi\pi}^0(k)$ is the bare susceptibility of magnon mode in Eq. (\ref{bs}).

\section{Polarization bubble}\label{BB}

To evaluate the loop integral of the polarization bubble, we use the Feynman parametrization 
\begin{equation}
\frac{1}{AB} = \int_0^1 \frac{du}{[uA+(1-u)B]^2},
\end{equation}
such that the loop integral in Eq. (\ref{Pi}) becomes 
\begin{equation}\label{fp}
\begin{split}
& \Pi _{\pi } (q) = \frac{m_0^4 r^2}{2} \int _0^1du\int _{\Lambda }\frac{d^Dk}{(2 \pi )^D}\frac{1}{\left[k^2+C(u,q)^2\right]^2}  \\
 &=\frac{m_0^4 r^2}{2} \int_0^1 du \int_0^\Lambda \frac{\rho^{D-1}d\rho}{(2\pi)^{D}} 
 \frac{\prod _{j=1}^{D-2}\int _0^{\pi } \sin ^j \theta_j d\theta _j
 \int_0^{2\pi} d\phi}{\left[\rho^2+C(u,q)^2\right]^2}   \\
 & =\frac{m_0^4 r^2}{2}
 \begin{cases}
\frac{1}{4\pi^2}\int_0^1du \left(\frac{1}{C} {\tan ^{-1}\left(\frac{\Lambda }{C}\right)}-\frac{\Lambda }{\Lambda^2 + C^2}\right), & D=3 \\
\frac{1}{16\pi^2}\int_0^1du \left(\log \left(\frac{\Lambda ^2+C^2}{C^2}\right)-\frac{\Lambda ^2}{\Lambda^2+C^2}\right), & D=4 
 \end{cases}
\end{split}
\end{equation}
where $C(u,q)=\sqrt{q^2 u (1-u)+A/2}$ and $d^Dk=\rho^{D-1}d\rho \prod _{j=1}^{D-2} \sin ^j \theta_j d\theta _j  d\phi$ in the spherical coordinate. For the ultraviolet cutoff $\Lambda \gg C$, the integrals in Eq. (\ref{fp}) reduce to Eq. (\ref{Pi}) that reproduces the results in Ref. \onlinecite{PodolskyArovasPRB} in the limit $A\rightarrow 0$.

\section{Cancellation of tadpole diagrams}\label{CC}

The full Higgs mode susceptibility is
\begin{equation}\label{tadpole}
\chi_{\sigma\sigma}(q) = \int d^Dx e^{-i q \cdot x}  \frac{\int \mathcal{D}\bm{\Phi}(x)  \sigma(x)\sigma(0)  e^{-\mathcal{S}} }{Z}.
\end{equation}
Employing the perturbation expansion to the first-order, we obtain
\begin{equation}
\begin{split}
\chi_{\sigma\sigma}^{(1)}(q)= &\int d^Dx e^{-i q \cdot x} \bigg( \frac{m_0^2}{2g}\int d^Dx' G_{\sigma\sigma}^{(0)}(x-x')  G_{\pi\pi}^{(0)}(0)   G_{\sigma\sigma}^{(0)}(x')  \\
&+\frac{(r^2-1)m_0^2}{2g} \int d^D x' G_{\sigma\sigma}^{(0)}(x-x')   G_{\sigma\sigma}^{(0)}(x') \bigg),
\end{split}
\end{equation}
where the first term is the tadpole diagram from $\mathcal{S}_A$ and the second  term comes from the counterterm $\mathcal{S}_C$. According to Eq. (\ref{couter}), we have $G_{\pi\pi}^{(0)}(0) = 1-r^2$ in the large $N$ limit such that the tadpole diagram in Eq. (\ref{tadpole}) is cancelled by the counterterm. Then the first order susceptibility vanishes $\chi_{\sigma\sigma}^{(1)}(q) = 0$.

\bibliography{references}

\end{document}